\documentclass[letterpaper,twocolumn,longbibliography,pre,amsmath,amssymb,float,notitlepage]{revtex4-2}

\usepackage{xcolor}
\usepackage{cancel}
\usepackage{graphicx}
\usepackage{dcolumn}
\usepackage{bm}

\begin{document}

\preprint{APS/123-QED}

\title{Integer Fluxonium Qubit
}

\author{Raymond A. Mencia$^{1,2}$}

\author{Wei-Ju Lin$^{2}$}

\author{Hyunheung Cho$^{2}$}

\author{Maxim G. Vavilov$^{3}$}

\author{Vladimir E. Manucharyan$^{1,2}$}
\affiliation{$^{1}$Institute of Physics, Ecole Polytechnique Federale de Lausanne, 1015 Lausanne, Switzerland}
\affiliation{$^{2}$Department of Physics, Joint Quantum Institute, and Quantum Materials Center, University of Maryland, College Park, MD 20742, USA}
\affiliation{$^{3}$Department of Physics, University of Wisconsin-Madison, Madison, WI 53562,  USA}

\date{\today}
\pacs{}
\begin{abstract}

We describe a superconducting qubit derived from operating a properly designed fluxonium circuit in a zero magnetic field. The qubit has a frequency of about 4 GHz and an energy relaxation quality factor $Q \approx 0.7\times 10^7$, even though the dielectric loss quality factor of the circuit components is in the low $10^5$ range. The Ramsey coherence time exceeds 100 $\mu$s, and the average fidelity of Clifford gates is benchmarked to $\mathcal{F} > 0.999$. These figures are expected to improve with optimized fabrication and measurement procedures. Our work establishes a ready-to-use ``partially protected" superconducting qubit functioning in the frequency range of conventional transmons.
\end{abstract}
                        
\maketitle

\section{\label{sec:level1} Introduction}

In recent years, superconducting fluxonium qubits have reached~\cite{nguyen2019high, zhang2021universal,ficheux2021fast, xiong2022arbitrary, bao2022fluxonium,moskalenko2022high,somoroff2024fluxonium} and may have even exceeded \cite{somoroff2023millisecond,ding2023high} the state-of-the-art coherence time and gate fidelity, defined by the industry-standard transmons~\cite{koch2007charge, schreier2008suppressing, paik2011observation}. 
Circuit-wise, the difference between the two devices seems minimal. A transmon is fundamentally a weakly anharmonic electromagnetic oscillator defined by a Josephson junction's inductance and a shunting capacitance \cite{koch2007charge}, while a fluxonium contains an additional high-inductance (superinductance) shunt, which nevertheless dramatically increases the qubit anharmonicity without introducing new decoherence channels \cite{manucharyan2009fluxonium, koch2009charging,manucharyan2012evidence}.  The strongly anharmonic spectrum of fluxoniums is an important additional resource for superconducting quantum processors, as it can help mitigate the propagation of coherent errors \cite{nguyen2022blueprint}. Another notable distinction is that fluxoniums operate at a much lower frequency, usually ranging around $100 - 1000$ MHz,  
as opposed to the typically 4-6 GHz frequency range for transmons. In fact, given that the dielectric loss \cite{martinis2005decoherence, wang2015surface, gambetta2016investigating, siddiqi2021engineering} is the primary decoherence mechanism in both devices, it is essentially the lower frequency of fluxoniums that allows matching the best transmons with far less sophisticated material science research and fabrication procedures \cite{yan2016flux, klimov2018fluctuations, place2021new, osman2021simplified, wang2022towards, kono2024mechanically}. 

Other alternatives to the transmon qubits are being actively explored. One direction is developing control techniques for even lower frequency fluxoniums \cite{earnest2018realization,zhang2021universal, najera2024high} which are expected to have even longer coherence time, albeit not necessarily higher quality factor. The more sophisticated ``protected" qubits can presumably have arbitrarily strong protection from any local decoherence source by operating at a nearly zero frequency~\cite{ioffe2002possible, kitaev2006protected, brooks2013protected}. However, these elegant ideas rely on a high degree of symmetry in the circuit Hamiltonians, which is challenging to implement with the present level of fabrication disorder \cite{kreikebaum2020improving,groszkowski2018coherence}. A more practical direction is exploring ``partially protected" qubits~\cite{gyenis2021moving}, for example, qubits in the transmon frequency range that are to some degree decoupled from dielectric loss, such as the recently demonstrated ``soft $0-\pi$"  \cite{gyenis2021experimental} and bifluxon qubits~\cite{kalashnikov2020bifluxon}. A significant effort is dedicated to hardware-efficient quantum error correction with bosonic codes, based on using transmons as switches rather than qubits, to control the quantum states of radiation in linear microwave cavities~\cite{Ma.2021, joshi2021quantum}. Such schemes could benefit from substituting a transmon with a lower-error rate control device.

Here, we circle back to conventional fluxoniums and point out that they can equally well operate as relatively high-frequency qubits when biased at the less explored integer flux quantum sweet spot, as opposed to the usual half-integer flux quantum sweet spot. Circuit parameters must be adjusted such that the low-energy dynamics is governed entirely by flux quantization in the loop and is well described by the dual of the Cooper pair box Hamiltonian \cite{bouchiat1998quantum}. In this case, the lowest excited state is a doublet, originating from the classical degeneracy of biasing an inductance $L$ with a positive vs. a negative flux quantum at energy $(h/2e)^2/2L$. Tunneling weakly splits the doublet. The lower-frequency transition from the ground state defines the integer fluxonium qubit, while the higher-frequency transition is forbidden by the parity selection rule. The doublet has been spectroscopically observed~\cite{manucharyan2009fluxonium} and characterized in the time domain~\cite{manucharyan2012evidence} in the very first fluxonium device. A more recently published study reports the observation of comparably long relaxation and coherence times, well in excess of $100$\,$\mu$s, at both sweet spots \cite{mencia2023ultra}. 
This work demonstrates that the doublet nature of the fluxonium transition at an integer flux quantum bias is compatible with precision single-qubit gates, the average fidelity of which is benchmarked to higher than $0.999$. Our result adds a ready-to-use qubit, nicknamed ``integer fluxonium", to the relatively short list of high-performance superconducting qubits.


\begin{figure*}
\centering
\includegraphics[width=14cm]{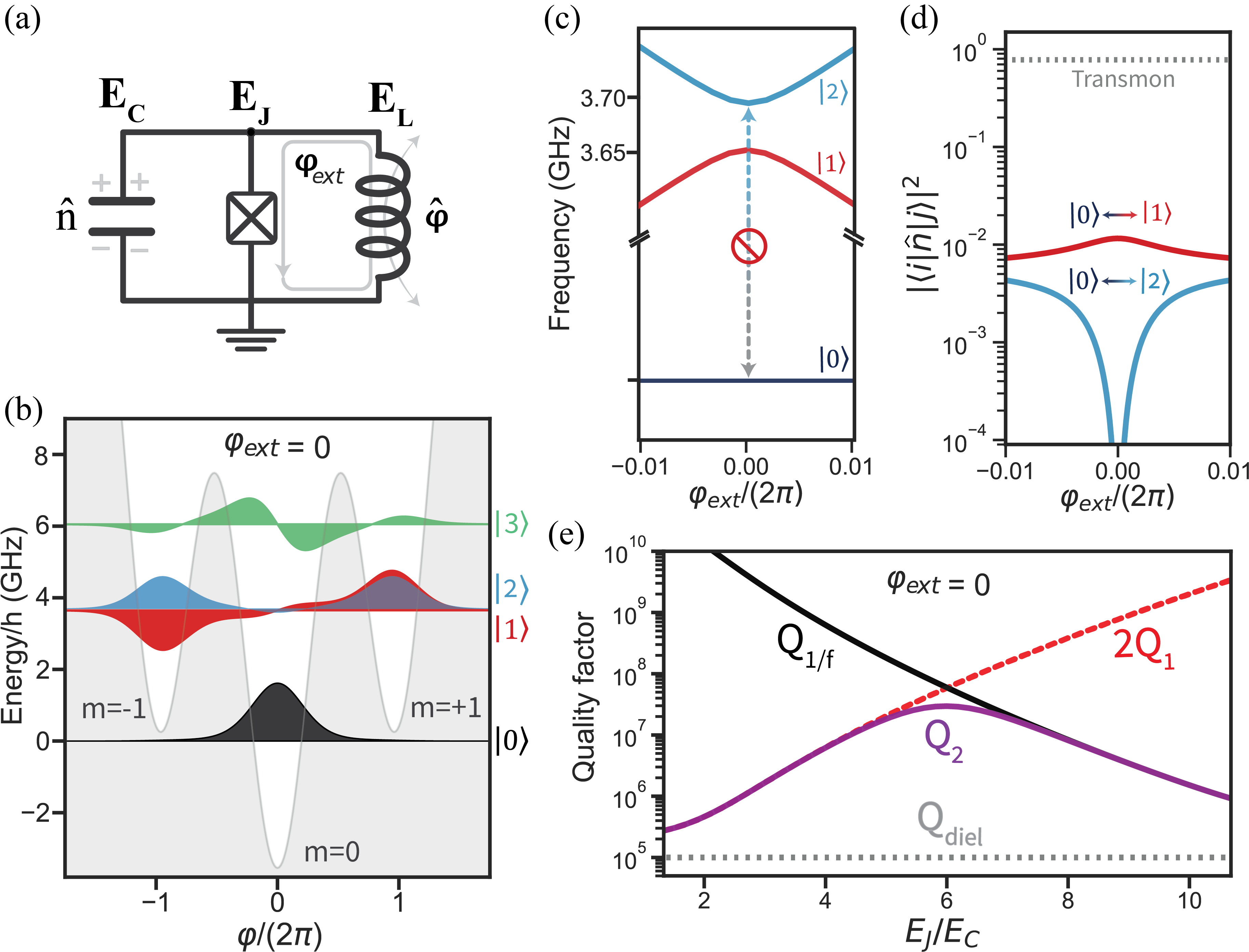}
\caption{ (a) Fluxonium circuit model (see text and Eq.~1). (b) An example of the effective potential energy profile for the phase degree of freedom $\hat \varphi$ in zero magnetic field ($\varphi_{\rm{ext}} =0$) superimposed on the lowest four energy levels and their wavefunctions. The integer fluxonium qubit transition is between states $|0\rangle$ and $|1\rangle$. (c,d) Frequencies and charge operator $\hat n$ matrix elements of the two lowest transitions calculated near zero field. (e) Illustration of balancing between energy relaxation quality factor $Q_1$ and $1/f$ flux noise dephasing quality factor $Q_{1/f}$ for finding optimal integer fluxonium parameters. Using example values of the $1/f$ flux noise amplitude $A_{1/f} = 1 \times 10^{-6}$ and the base dielectric loss quality factor $Q_{\textrm{diel}} = 1 \times 10^{5}$ (see section II-B), there is an optimal value $E_J/E_C \approx 6$ that maximizes the qubit's total decoherence quality factor $Q_2 = (2Q_1^{-1} + Q_{1/f}^{-1})^{-1}$. The simulation parameters are given in Table 1. Note, the qubit frequency $\omega_{01}$ is nearly independent on $E_J$ and $E_C$ and is approximately fixed by $\hbar \omega_{01} \approx 2\pi^2 E_L$.
}
\label{Fig1}
\end{figure*}


The paper is organized as follows. In section II we demonstrate the equivalence of the integer and the half-integer flux sweet spots with respect to the energy relaxation rate using an elementary model of flux quantization in a superconducting loop. In section III we describe devices and data, including a complete case study for device D in Table 1. In section IV we discuss the results and provide concluding remarks.

\section{Theory of integer fluxonium}

The fluxonium circuit (see Fig.~1a) is defined by three parameters: the Josephson energy $E_J$ of the Josephson junction, the charging energy $E_C = e^2/2C$ of the total capacitance $C$ and the inductive energy $E_L = (\hbar/2e)^2/L$ of the superinductance $L$. The junction and the inductance form a superconducting loop, which can be biased by an external magnetic flux $\varphi_{\textrm{ext}}\times (\hbar/2e)$. Quantum mechanics of fluxoniums is usually described by a pair of conjugate macroscopic variables, the superconducting phase-difference $\varphi$ across the Josephson junction and the charge $2e\times n$. The two observables, the phase, $\hat\varphi$, and the Cooper pair number operator, $\hat{n}$, displaced at the capacitor satisfy the standard position-momentum-like commutation relation $[\hat{\varphi}, \hat{n}] = i$ and the Hamiltonian reads 
\begin{equation}
H/h = 4E_C \hat{n}^2  - E_J \cos(\hat{\varphi} - \varphi_{\rm{ext}}) +\frac12 E_L \hat{\varphi}^2.
\label{daddy}
\end{equation}
The parameter regime for integer fluxonium is derived from the following arguments. First, as with all fluxoniums, it is essential to have $E_L \ll E_J$, in which case the potential energy (the $\varphi$-dependent term of Eq.~(1)) has multiple local minima  (Fig.~\ref{Fig1}b). The charging energy term is equivalent to the kinetic energy of a particle with a mass $C$ moving in such a potential. There are two types of excitations: oscillations in a single well (plasmon) and tunneling between the wells (fluxon). One must decouple these excitations to implement integer fluxonium, which is achieved by setting the plasmon frequency sufficiently above the fluxon frequency. In the asymptotic limit of weak tunneling, that is $E_J/E_C \gg 1$, the plasmon energy is given approximately by $\sqrt{8 E_J E_C}$ while the fluxon energy is approximately $2\pi^2 E_L$, which reduces the parameter requirement for the integer fluxonium to $\sqrt{8E_J E_C} \gg 2\pi^2E_L$, with qubit frequency given approximately by $\hbar\omega_{01} \approx 2\pi^2E_L$, the offset between the lowest two potential wells in Fig.~1b. Similarly to conventional fluxoniums, the far-detuned plasmon transition enables a dispersive qubit-cavity interaction used for the qubit readout, as long as the cavity frequency is chosen to be near the plasmon transition frequency ~\cite{zhu2013circuit}.

\begin{table*}[t]

\begin{tabular}{|c|l|c|c|c|c|c|}

\hline 

\textrm{}&
\textrm{Device}&
\textrm{A}&
\textrm{B}&
\textrm{C}& 
\textrm{D}&
\textrm{Fig. 1}\\
\hline
 
\noindent\textbf{Fluxonium model fit}

\textrm{}&$E_J$ (GHz) & 4.12 & 3.84 & 7.20 & 6.78 & 5.0\\
\textrm{}&$E_C$ (GHz) & 1.64 & 1.75 & 2.04 & 1.47 & 1.5\\
\textrm{}&$E_L$ (GHz) & 0.18 & 0.14 & 0.18 & 0.22 & 0.2\\

\hline
\noindent\textbf{Flux tunneling model fit}

\textrm{}&$E_L^{\Sigma}$ (GHz) & 0.163 &  0.128 & 0.175 &  0.206 &-\\
\textrm{}&$\epsilon_{1}$ (MHz) & 308 & 362 & 190 & 91&-\\
\textrm{}&$\epsilon_{2}$ (MHz) & 73 & 81 & 6 & 9&-\\
\hline

\textbf{Measured parameters}

\textrm{}&$\omega_{01}/2\pi$ (GHz) & 3.20 & 2.51 & 3.45 & 4.14&-\\
\textrm{}&$\omega_{12}/2\pi$ (MHz) & 103 & 106 & 24 & 11&-\\

\textrm{}&$\Bar{T}_1$ ($\mu$s) & 109  & 101 & 328 & 255&-\\
\textrm{}&$\Bar{T}_2^E$ ($\mu$s)& 175 & 201 & 81 & 185&-\\
\textrm{}&$\Bar{T}_2^*$ ($\mu$s)& 38 & 61 & 57 & 118&- \\
\hline

\textbf{Inferred parameters}

\textrm{}&$Q_{\text{diel}}$ & $3.5 \times 10^5$&$3.6 \times 10^5$ & $2.6 \times 10^5$ & $1.0 \times 10^5$&$1.0\times 10^5$ \\ 
\textrm{}&$|n_{01}|^2$ (IFQ) & $2.1 \times 10^{-2}$ & $1.8 \times 10^{-2}$ & $4.4 \times 10^{-3}$ & $3.1 \times 10^{-3}$& $1.2 \times 10^{-2}$\\
\textrm{}&$|n_{03}|^2$ (IFQ) & $1.6 \times 10^{-1}$ & $1.4 \times 10^{-1}$ &$2.2 \times 10^{-1}$ & $2.9\times 10^{-1}$&$2.2 \times 10^{-1}$\\
\textrm{}&$|\chi_{01}/\chi_{02}|$ & {3.0} &{0.7}  & {0.1}&{8.4} &-\\
\hline  
\end{tabular}
\caption{\label{tab:table2}
A summary of device parameters.All ``measured parameters" imply $\varphi_{\textrm{ext}} = 0$.
}
\end{table*}

\subsection{Flux-dual Cooper pair box model}

Although spectral properties of integer fluxonium can be calculated numerically from the Hamiltonian (\ref{daddy}) for arbitrary circuit parameters and flux bias (see Fig.~\ref{Fig1}c,d), it is insightful to consider an effective model defined entirely in terms of flux quantization and tunneling \cite{matveev2002persistent, koch2009charging}. We start with the basis of fluxon states $|m\rangle$ corresponding to a particle resting at the bottom of the m-th Josephson well (in reality, we are only interested in the lowest three wells shown in Fig.~\ref{Fig1}a). Introducing the phenomenological tunneling energy amplitudes $\epsilon_1$ to change $m$ by $\pm1$ and $\epsilon_2$ to change $m$ by $\pm 2$ we can write the following effective Hamiltonian 
\begin{equation}
\begin{aligned}
H_{\textrm{eff}}/h = \sum_m \frac{E_L^{\Sigma}}{2} (2\pi m-\varphi_{\textrm{ext}})^2  |m\rangle \langle m|
+\\
 -  \frac{\epsilon_1}{2}\sum_m  |m\rangle \langle m\pm 1| +\frac{\epsilon_2}{2}\sum_m|m\rangle \langle m \pm2|,
    \end{aligned}
    \label{baby}
\end{equation}
where $E_L^{\Sigma} = (E_L^{-1}+E_J^{-1})^{-1}$ and is the total linearized loop inductance. The values of the tunneling amplitudes $\epsilon_1$ and $\epsilon_2 \ll \epsilon_1$ depend on the three circuit parameters $E_J$, $E_C$, and $E_L$, but it is safe to assume that both amplitudes are exponentially suppressed with the parameter $\sqrt{8E_J/E_C}$. The double-fluxon tunneling amplitude $\epsilon_2$ is not necessary to derive the main properties of integer fluxonium. Still, it is necessary to obtain an accurate value of $\epsilon_1$ by matching the low-energy spectra of the two Hamiltonians (1,2). In fact, for $\epsilon_2=0$, the Hamiltonian (2) describes a Cooper pair box (a charge qubit) in the regime of strong charge quantization, where charging energy is replaced by the inductive energy, Cooper pairs with fluxons, and offset charge with the flux bias.

The spectrum of the model Hamiltonian (\ref{baby}) near zero bias can be obtained by diagonalizing the corresponding $3\times3$ matrix in the basis of three fluxon states $|m=-1\rangle$, $|m=0\rangle$, and $|m=+1\rangle$. Setting $\varphi_{\textrm{ext}}=0$ in Eq.~\ref{baby} and introducing a small parameter $\alpha = \epsilon_1/4\pi^2 E_L^{\Sigma}$ we get the following perturbative expressions for the three lowest-energy eigenstates:
\begin{equation}
    \begin{aligned}
        |0\rangle & = |m=0\rangle + \alpha \big(|m=-1\rangle + |m=+1 \rangle\big),
        \\
         |1\rangle & = \frac{1}{\sqrt{2}} \big( |m=-1\rangle -|m=+1 \rangle\big),
         \\
        |2\rangle & = \frac{1}{\sqrt{2}}\big(|m=-1\rangle + |m=+1 \rangle)  - \sqrt{2} \alpha  |m=0\rangle .
    \end{aligned}
\end{equation}
The eigenstates $|1\rangle$ and $|2\rangle$ define the excited-state doublet of integer fluxonium. The first two transition frequencies are given by 
$\hbar \omega_{01} = 2\pi^2 E_L^{\Sigma} + \alpha \epsilon_1 - \epsilon_2/2$ and $\hbar\omega_{02} = 2\pi^2 E_L^{\Sigma} + 2\alpha \epsilon_1 + \epsilon_2/2$. Thus, the doublet is split by $\hbar\omega_{12}  = \alpha\epsilon_1
+ \epsilon_2$. 
Note that $\epsilon_2$ parameter does not affect the eigenstates but merely increases the doublet splitting $\omega_{12}$. In the qualitative analysis below, we will set $\epsilon_2$ to zero to simplify the expressions without losing the essence.

The matrix elements of $\hat{\varphi}$ and $\hat{n}$ operators can be obtained by noting the relation $\hat{\varphi} = 2\pi \hat{m}$ and using the identity $\langle i|\hat{n}|j\rangle = \langle i|\hat{\varphi}|j\rangle \times (\hbar\omega_{ij}/8E_C$), which follows directly from the commutation rule and Eq.~\eqref{daddy}. For the phase-diference matrix elements, we get $\langle 0|\hat{\varphi}|2\rangle =0$ and $\langle 0|\hat{\varphi}|1\rangle \approx 2\pi\alpha \ll 1$. Note, in case of a conventional fluxonium operation at $\varphi_{\textrm{ext}}=\pi$, within the model of Eq.~(\ref{baby}), we have  $\hbar \omega_{01} = \epsilon_1$ and $\langle 0|\hat{\varphi}|1\rangle \approx \pi$, which is a much larger number than at the integer flux bias. Yet, the matrix elements of the conjugate Cooper pair number operator $\hat{n}$ come out essentially the same at the two sweet spots: 
\begin{equation}
\langle 0|\hat{n}|1\rangle_{\varphi_{\textrm{ext}} =0} \approx \sqrt{2}\langle 0|\hat{n}|1\rangle _{\varphi_{\textrm{ext}} =\pi} \approx \frac{\sqrt{2}\pi}{8} \frac{\epsilon_1}{E_C}.
\end{equation}
The degree of protection of integer fluxonium from energy relaxation can be judged by comparing this matrix element estimate with the typical transmon value, $\sqrt[4]{E_J/8E_C} \simeq 1$ \cite{krantz2019quantum}.

The above qualitative analysis based on Eq.~\eqref{baby} compliments accurate numerical calculations based on Eq.~\eqref{daddy}. Examples of calculated wave functions, transition frequencies, and matrix elements are shown in Fig.~(\ref{Fig1}b-d). 

\subsection{Dielectric loss vs. $1/f$ flux noise}

An optimal design for an integer fluxonium must simultaneously suppress the effect of the two prevailing decoherence mechanisms: the dielectric loss and the $1/f$ flux noise. Both channels originate from defects in the surface oxide layers of superconducting leads and thus cannot be readily eliminated, at least not in the near future. Let us define the dielectric loss via the effective loss tangent $\tan\delta_C$ of the circuit's total capacitance $C$~\cite{Martinis.2005,Megrant.2012,Wang.2015}. For both transmon and fluxonium qubits, the energy relaxation quality factor $Q_1$ is related to $Q_{\textrm{diel}}$ according to the expression of Fermi's Golden rule  $Q_1 = \omega_{01} T_1$, and $1/T_1 =   32\pi E_C|\langle 0|\hat{n}|1\rangle|^2/Q_{\textrm{diel}}$, where we define $Q_{\textrm{diel}} = 1/\tan\delta_C$ as the effective dielectric loss quality factor. For a transmon, $Q_1 = Q_{\textrm{diel}}$. For the best transmons,  $Q_{\textrm{diel}} \approx  (0.1 -1) \times 10^7$~\cite{Kamal.2016.arXiv,nersisyan2019manufacturing,Place.2021,gordon2022environmental}. For integer fluxonium, we can qualitatively estimate the parameter dependence of $Q_1$ as
\begin{equation}
\label{Q1}
Q_1\propto \frac{E_C}{\hbar\omega_{12}} \times Q_{\textrm{diel}} \,.
\end{equation}
To estimate the dephasing rate due to the $1/f$ flux noise, we define the noise spectral density as $S_{1/f}(\omega) = 2\pi A_{1/f}^2 (h/2e)^2/\omega$, where the quantity $A_{1/f}$ characterizes the strength of the flux noise. Slow variation in the value of $\varphi_{\textrm{ext}}$ induces a slow variation of the qubit frequency to second order in the value of $A_{1/f}$ and proportionally to the curvature of the qubit transition with respect to the flux bias. We estimate the resulting dephasing time as $T_{1/f}^{-1} \approx A_{1/f}^2 \times \rm \partial^2\omega_{01}/\partial(\varphi_{\rm{ext}}/2\pi)^2$
\cite{ithier2005decoherence}. The resulting dephasing quality factor $Q_{1/f} = \omega_{01}T_{1/f}$ is given up to a numerical factor:
\begin{equation}
Q_{1/f} \propto \frac{\hbar\omega_{12}}{E_L} \times \frac{1}{A_{1/f}^2} \,.
\end{equation}
Reducing the frequency $\omega_{12}$ would increase $Q_1$ but reduce $Q_{1/f}$ by the same amount, without changing the qubit transition frequency $\hbar\omega_{01} \approx 2\pi^2 E_L$. 
The combined decoherence quality factor $Q_2 = 1/(1/2Q_1 + 1/Q_{1/f})$ reaches its maximum for a specific value of $E_J/E_C$, which depends on the experimental values of $Q_{\textrm{diel}}$ and $A_{1/f}$. In other words, the strength of flux noise $A_{1/f}$ determines how much the qubit's quality factor can be increased over the dielectric loss limit $Q_{\textrm{diel}}$. Fig.~1e illustrates the described balance using experimentally motivated values of $A_{1/f} = 10^{-6}$ and $Q_{\textrm{diel}} = 10^{5}$, resulting in an enhancement of the total coherence quality factor $Q_2$, that is $Q_2/Q_{\textrm{diel}}\gg 10$. 

\subsection{Leakage to state $|2\rangle$}

The more integer fluxonium is made protected from energy relaxation, the smaller the splitting between the computational level $|1\rangle$ and the non-computational level $|2\rangle$, see Eq.~\eqref{Q1}. As long as $\varphi_{\textrm{ext}} =0$, no leakage to state $|2\rangle$ is expected during qubit manipulations. Indeed, a resonant qubit drive at a frequency around $\omega_{01}$ does not couple to transition $|0\rangle$-$|2\rangle$ because $\langle 0|\hat{n}|2\rangle =0$, nor does it couple to transition $|1\rangle$-$|2\rangle$ because of the large frequency detuning. Thus, in theory, integer fluxonium is effectively a 2-level system: even transition $|0\rangle$-$|3\rangle$ would not be coupled by the drive because of its large detuning.

In practice, though, there is always a small deviation $\delta\varphi_{\textrm{ext}} \ll 1$ of the value $\varphi_{\textrm{ext}}$ from an integer value, either due to an instrument imprecision, or a background field, or at the very least due to the same $1/f$ flux noise that causes pure dephasing. Consequently, the qubit drive acquires a small coupling to state $|2\rangle$, proportional to the value of $(d \langle 0|\hat{n}|2\rangle/d\varphi_{\textrm{ext}})\delta \varphi_{\textrm{ext}}$ near $\varphi_{\textrm{ext}}=0$. The frequencies $\omega_{01}$ and $\omega_{02}$ would also be slightly shifted to the second order in $\delta\varphi_{\textrm{ext}}$. For a given $\delta\varphi_{\textrm{ext}}$, we can model the resulting coherent errors in our qubit using the following elementary $N+1$-level Hamiltonian
\begin{equation}
\begin{aligned}
\label{Hdrive}
H_{\textrm{drive}}(\delta\varphi_{\textrm{ext}})/\hbar =\sum_i^N \omega_{0i}|i\rangle\langle i| 
+\\
+\hat{n} \big(I(t)\cos\omega t + Q(t)\sin \omega t\big),
\end{aligned}
\end{equation}
where $\omega_{0i} = (E_i-E_0)/\hbar$ is the transition frequency for eigenenergy states $|0\rangle$ and $|i\rangle$ of Eq.~\eqref{daddy} detuned from the sweet spot by $\delta \varphi_{\textrm{ext}}$, and $\hat n$ is the Cooper pair number operator in the eigenstate basis of Eq.~\eqref{daddy}. For the microwave drive we optimize $\omega \approx \omega_{01}$ and $I(t)$ and $Q(t)$ are slowly varying envelopes defining the qubit pulse shape. We use a numerical solution of this Hamiltonian to interpret the results of randomized benchmarking of experimental devices. We used $N=5$ lowest fluxonium eigenstates in simulations below.

\begin{figure}[t]
\centering
\includegraphics[width=8.5cm]{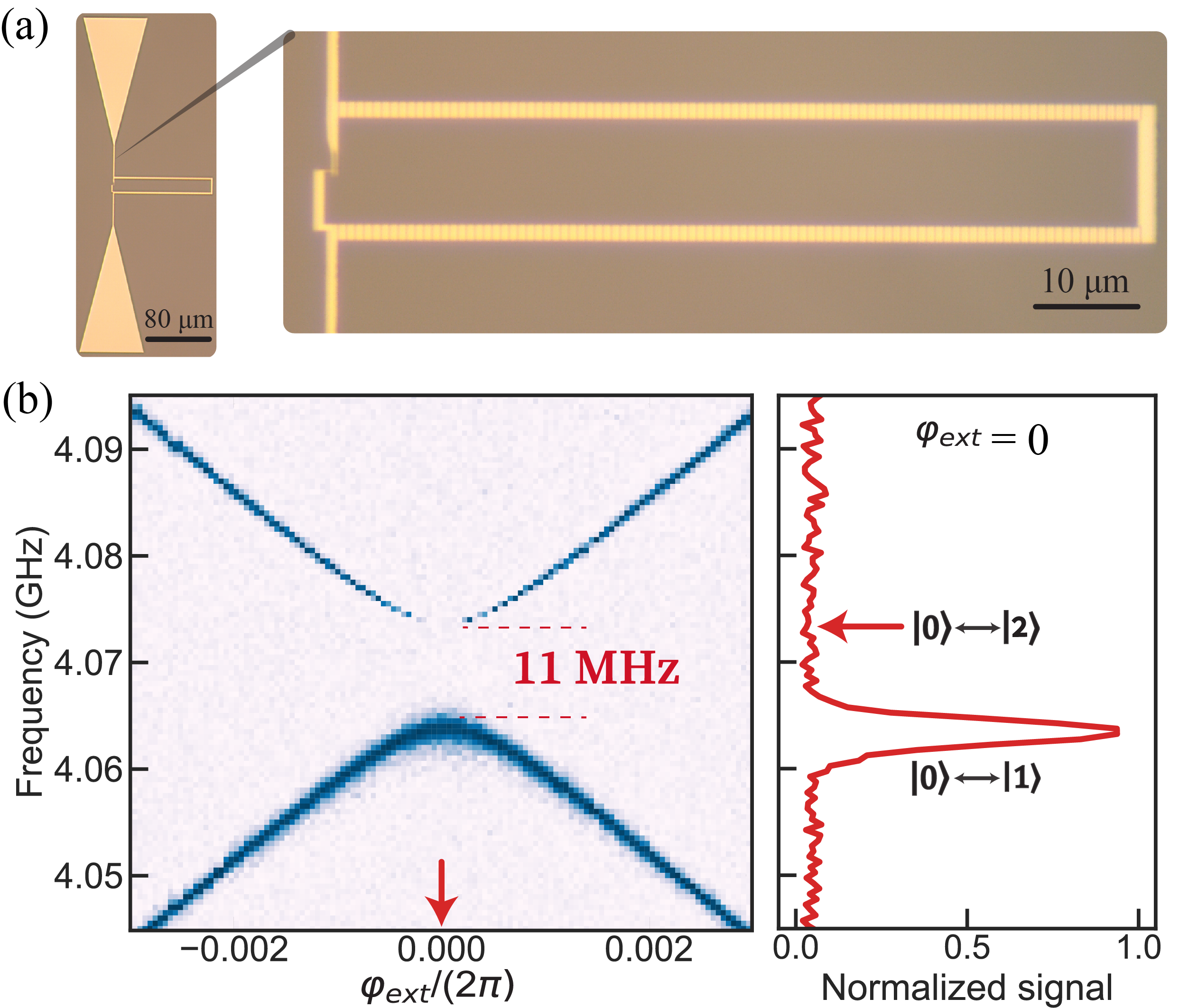}
\caption{(a) Optical image of an integer fluxonium device. 
(b) Two-tone spectroscopy signal (arbitrary units) as a function of frequency and flux bias near $\varphi_{\textrm{ext}}=0$. Inset shows a cross-section at $\varphi_{\textrm{ext}}=0$. Note the absence of the $|0\rangle$-$|2\rangle$ within the given signal-to-noise ratio. 
}
\label{exp_device_spect}
\end{figure}

\section{Qubit Characterization}\label{sec2}

\subsection{Spectroscopy and decoherence}
Spectroscopy and time-domain data were collected on four different devices A, B, C, and D. The device parameters are extracted from fitting the model in Eqs.~\eqref{daddy} and \eqref{baby} and are summarized in Table 1. In comparison to Ref. \cite{nguyen2019high}, we increased the charging energy $E_C$ and the Josephson energy $E_J$ but reduced the inductive energy $E_L$. Similarly to Ref. \cite{somoroff2023millisecond}, the readout is performed using a capacitive coupling to a 3D cavity for devices A, B, and D. Device C was coupled inductively to an on-chip resonator and read out using a wireless coupling to a 3D waveguide, similarly to Ref. \cite{smith2016quantization, kou2018simultaneous,pechenezhskiy2020superconducting}. Fabrication and measurement procedures are similar to those in Ref. \cite{nguyen2019high}. We focus our presentation on device D as it is the most characteristic of the integer fluxonium regime.

Figure~2a shows a representative optical image of device A (similar to device D). Fig.~2b shows the spectrum of device D near $\varphi_{\textrm{ext}}=0$. The data reveals a doublet analogous to the one first observed in Refs. \cite{manucharyan2009fluxonium, pop2014coherent}, except the splitting at  $\varphi_{\textrm{ext}}=0$ is reduced to a mere 11 MHz. The inhibition of transition $|0\rangle$-$|2\rangle$ is illustrated in the inset of Fig.~1b. The next transition $|0\rangle$-$|3\rangle$ in device D is far-detuned at 6.86 GHz. The calculated Cooper pair number transition matrix element is $\langle 0|\hat{n}|1\rangle \approx 0.056$ and is at least an order of magnitude below the typical transmon value.

\begin{figure}[t]
\includegraphics[width=8.5cm]{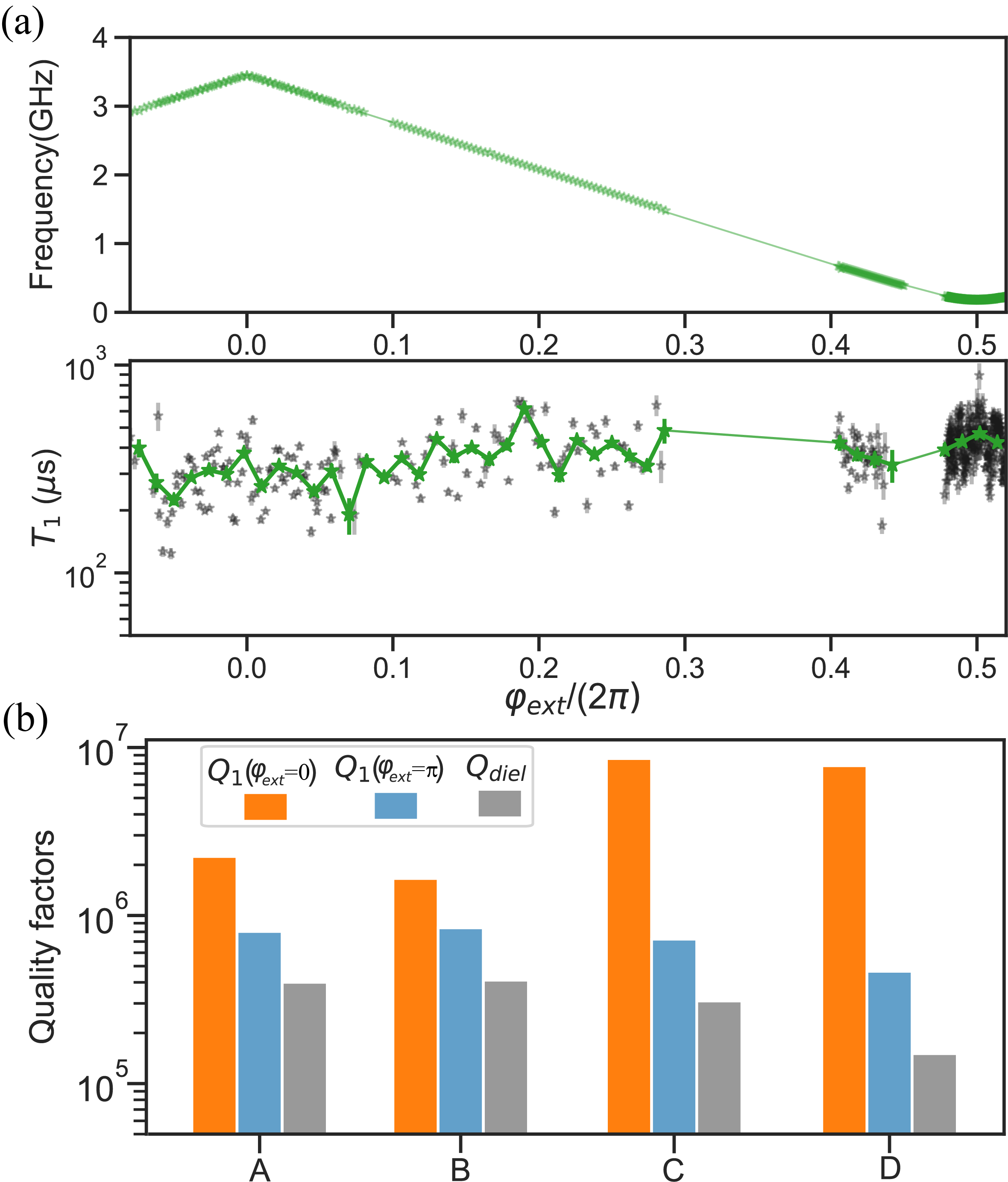}
\caption{ (a) The experimentally measured transition frequencies (top) and energy relaxation times (bottom) of device C as a function of external flux bias. The average values of $T_1$ are shown in green. (b) The quality factor of the measured qubit transitions due to energy relaxation, $Q_1$, at each flux sweet spot along with the extracted value of $Q_{\text{diel}}$. 
}
\label{relaxation}
\end{figure}

\begin{figure}[t]
\includegraphics[width=8.1cm]{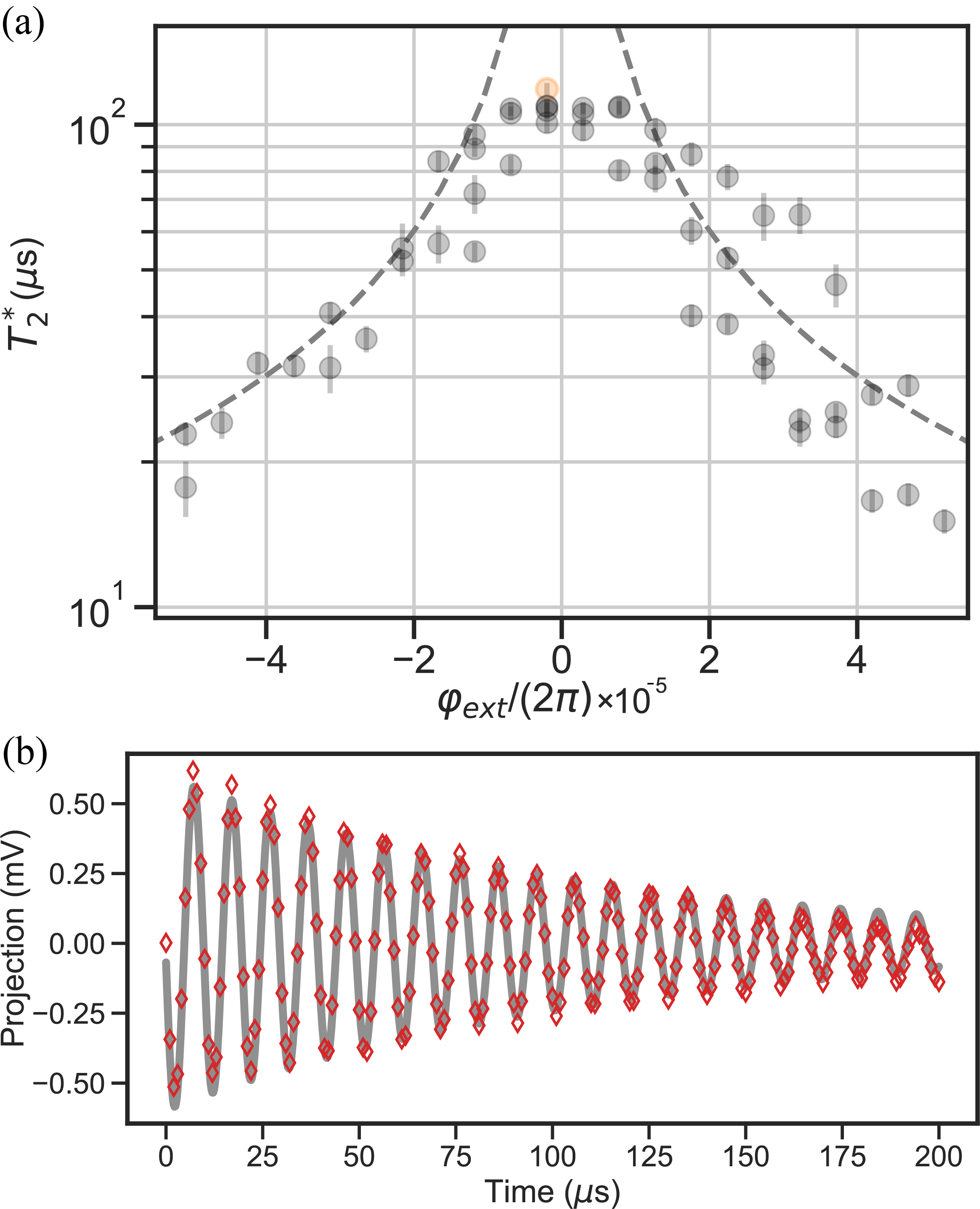}
\caption{ (a) The Ramsey coherence time $T_2^*$ of device D as a function of external flux. Dashed line is the first-order theoretical value $1/T_2^* = A_{1/f} |\partial \omega_{01}/\partial (\varphi_\text{{ext}}/2 \pi)| $ with $A_{1/f} \approx 8.8 \times 10^{-6}$.
(b) Example Ramsey fringe measured at $\varphi_{\textrm{ext}} =0$; corresponding to the orange data point in (a).}
\label{Ramsey}
\end{figure}

Figure~3 illustrates the frequency independence of the energy relaxation time $T_1$ over the range of several octaves in device C. Similar behavior was observed using less data points in all the other devices (see Fig.~7 in the Appendix). The relaxation data in Fig.~3a and Fig.~7 is well described by the dielectric loss model with a frequency-independent dielectric quality factor $Q_{\rm{diel}}$ (see discussion around Eq.~5). The qubit quality factor $Q_1 = \omega_{01} T_1$ in the more ``protected" designs C and D reach the values of $Q_1\simeq (6.5-7.0) \times 10^{6}$ range, which are comparable to those of optimally fabricated transmons~\cite{kono2024mechanically}. Nevertheless, the extracted dielectric quality factor $Q_{\rm{diel}}$ is only in the low $10^5$ range, which demonstrates the protection of integer fluxoniums from dielectric loss at work. Our experiment cannot exclude other energy relaxation channels, so the estimation of the $Q_{\textrm{diel}}$ should be taken as an upper bound. This low estimate is not unexpected, though, since our fabrication procedures did not involve any advanced tricks for mitigating the surface loss.

Interleaved measurements of $T_1$ and $T_2^E$ (the single $\pi$-pulse echo protocol) were done extensively over many hours to probe the repeatability of each characteristic time. We obtain the respective average times shown in Table 1 as $\bar{T}_1$ and  $\bar{T}_2^E$ by averaging each individual fit of the respective measurements. The coherence in devices A and B was limited by energy relaxation, as indicated by $T_2^E \approx 2T_1 \approx 200$ $\mu$s. In case of device D we describe a more accurate measurement of the Ramsey coherence time $T_2^*$ in the vicinity of the sweet spot $\varphi_{\textrm{ext}}=0$ (see Fig.~4).

At $\varphi_{\textrm{ext}}=0$ we find $T_2^* \approx 118~\mu\rm{s}$ and $T_2 ^{E}\approx 185~\mu\rm{s}$, which suggests the presence of low-frequency flux noise in the system. This noise can come from the conventional $1/f$ flux noise in superconducting devices, as well as even slower flux drifts due to imperfect shielding, which may occur at the time scale of data acquisition.  Measurement of $T_2^*$ slightly away from the sweet spot (Fig.~4a) agrees with the first-order flux-noise estimate $A_{1/f} = 8.8 \times 10^{-6}$. This estimate is about factor of 3 higher than what is usually found in fluxonium devices obtained using $T_2^{E}$ rather than the more conservative $T_2^*$ data.
The value of $T_2^E$ near the sweet spot can also be explained by the presence of thermally excited photons in the readout cavity at an effective temperature of $T = 50~\rm{mK}$. The estimate is made using the relations $1/T_2^E = \bar{n}\kappa\chi_{01}^2/(\kappa^2 + \chi_{01}^2)$ and $\bar{n} = \exp(-\hbar\omega_{cav}/k_B T)$, as well as the cavity frequency $\omega_{\rm{cav}}/2\pi=7.46 ~\rm{GHz}$, cavity total linewidth $\kappa = 2\pi \times 15~\rm{MHz}$ and the dispersive shift $\chi_{01} = 2\pi \times 14.3~\rm{MHz}$. It is unlikely that our actual cavity temperature is much lower, as the above value is typically found in well-filtered experiments.

\subsection{Gates benchmarking}

\begin{figure*}[t]
\centering
\includegraphics[width=18.0cm]{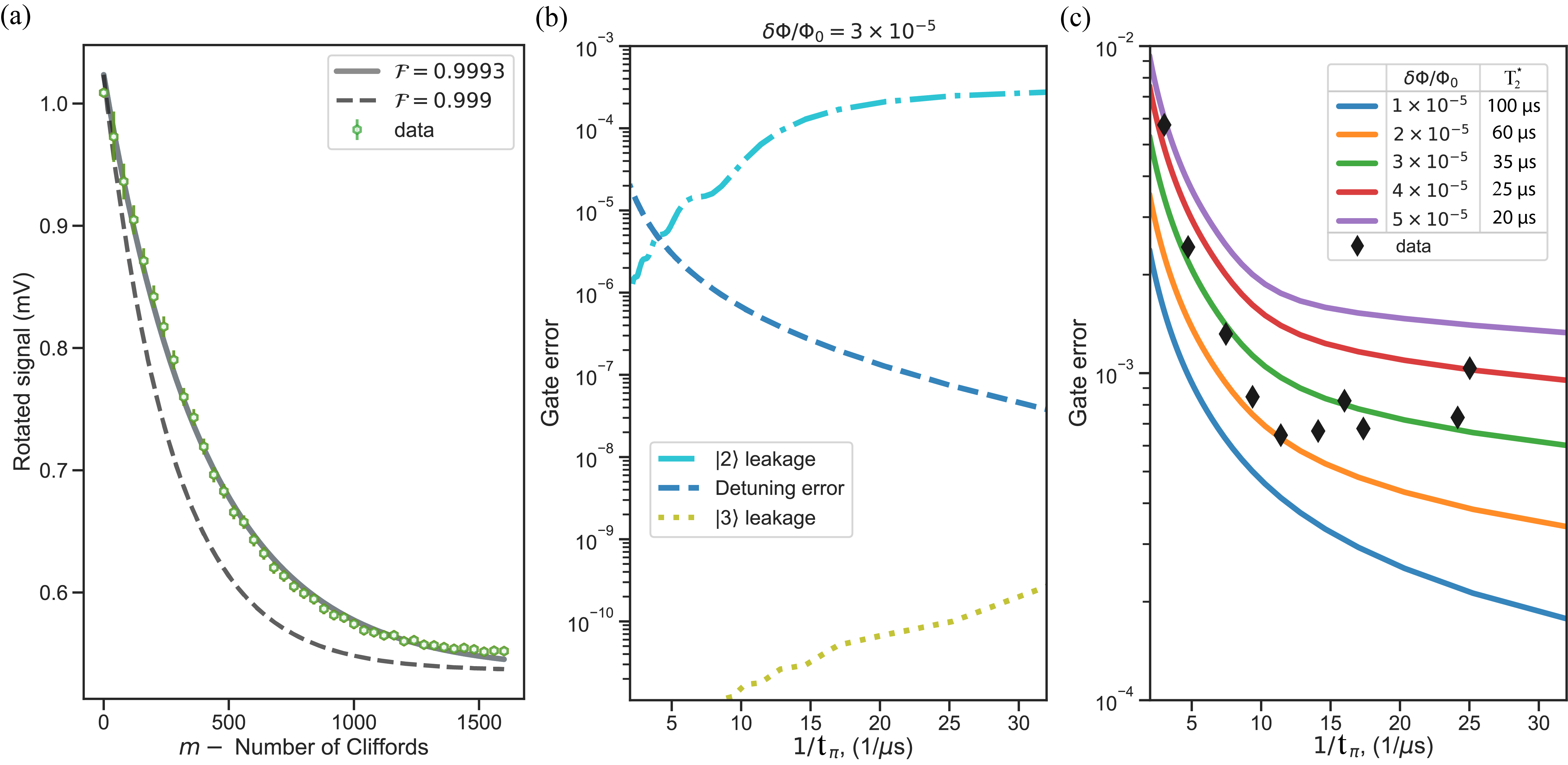}
\caption{(a) The RB trace for highest fidelity $\mathcal{F}=0.9993$, measured with a pi-pulse time $t_{\pi} = 88$ ns. The reference dashed line corresponds to a fidelity $\mathcal{F} = 0.999$, using the same A-B parameters. (b) Summary of calculated coherent errors: the detuning error (dashed line), the state $|2\rangle$ leakage (dot-dashed line), and the state $|3\rangle$ leakage (dotted line) at a flux offset of  $\delta\varphi_{\text{ext}}/2\pi= 3 \times 10^{-5}$. (c) RB measurements as a function of gate time $t_{\pi}$ (black diamonds) superimposed with the total gate errors (coherent and incoherent) calculated at several flux offset values. The incoherent error contribution was estimated using the values of $T_2^*$ from Fig.~4a.}
\label{RB}
\end{figure*}

We arrive at the most important demonstration of the potential usefulness of integer fluxonium qubit, benchmarking of the single-qubit Clifford-group gates. This test is especially critical due to the possibility of the leakage of quantum information outside the computational subspace of states $|0\rangle$ and $|1\rangle$. For benchmarking, we select device D as the most protected and, hence, most sensitive to the leakage.

The experimentally performed RB sequence consists of randomly chosen Clifford gates applied to the qubit before applying a single inversion gate; bringing the state back to the initial. The number of Clifford gates before the final state inversion gate is increased, resulting in a decaying probability of properly recovering the initial state. Measurements are fit according to the polynomial $A +B p^m$; where p is the depolarization parameter, $m$ is the number of Clifford gates and A, B are constants that account for state preparation and measurement (SPAM) errors \cite{lazuar2023calibration}. The average error rate $1-\mathcal{F}$ for Clifford group gates is defined as $(1-\mathcal{F}_{\textrm{Cliff}})=(1-p)/2$. Since there are on average 1.833 physical gates per Clifford the average physical gate fidelity is $(1-\mathcal{F})= (1-\mathcal{F}_{\textrm{Cliff}})/1.833$ \cite{somoroff2023millisecond,somoroff2024fluxonium}. The qubit pulses used where constant plateau-pulses with Gaussian edges that used a 5 ns ramp time.
The highest average physical gate fidelity measured was $\mathcal{F} = 0.9993(1)$ realized with a total pi-pulse time of $t_{\pi} = 88$ ns (see Fig.~5a). Note, the very small discrepancy of the fitted data from the exponential function can be due to fluctuations of the fidelity value in time during repetitive randomized benchmarking measurements. 
\\

\subsection{Leading  errors}

Randomized benchmarking measurements for device D are simulated in QuTip  \cite{johansson2012qutip} to quantify different contributing sources of error. The simulation uses 5 fluxonium eigenstates, and the same pulses as in the experiment. For the sake of simplicity, we ignore decoherence processes and consider their contributions separately. The theoretical estimate of the fidelity is obtained from Ref.~\cite{Pedersen2007}: 
$$
\mathcal{F}= \frac{\text{Tr}(U^{\dagger}_{\rm comp}U^{}_{\rm comp}) + |\text{Tr}(U_{\rm comp}^{\dagger}U^{}_{\text{ideal}}))|^2}{6},
$$ 
where $U_{\rm comp}$ is the evolution operator projected to the computational subspace.

In theory, the qubit drive induces no error due to the leakage to state $|2\rangle$ if we assume that the qubit circuit is biased strictly at $\varphi_{\textrm{ext}} =0$. The leakage to state $|3\rangle$ is also negligible in a reasonable range of $\pi$-pulse times $t_{\pi}$ (see Fig.~5b), thanks to the relatively large difference between frequencies $\omega_{01}$ and $\omega_{03}$ (see Table 1). A more likely scenario for errors is the occurrence of a small flux offset $\delta\varphi_{\textrm{ext}}$ away from the zero (integer) value. Such offsets can lead to three types of errors. (i) The parity selection rule is weakly broken such that $\langle 0| n|2\rangle \propto \delta\varphi_{\textrm{ext}}$, and hence a sufficiently strong qubit drive would introduce a leakage to state $|2\rangle$. (ii) If the flux-offset developed unnoticed after the gate pulse was calibrated at $\varphi_{\rm{ext}}=0$, the qubit frequency would shift by $\delta \omega_{01}$ and hence the qubit rotation would become imperfect. Consequently, there is a coherent ``detuning" error, on the order of $ (\delta \omega_{01}/2\pi)^2 t_{\pi}^2$ for the gate time $t_{\pi}$. (iii) Finally, a flux offset increases the sensitivity of the qubit frequency to the $1/f$ flux noise and hence reduces the coherence time $T_2^* = T_2^*(\varphi_{\rm{ext}})$ (see Fig.~4a). Consequently, we get a higher average incoherent error. This error is estimated for RB sequences as $(t_{\pi}/T_2^*(\varphi_{\rm{ext}}))/3$ in the limit $2 T_1 \gg T_2^*$~\cite{o2015qubit,chen2018metrology}. We consider all three types of errors in our analysis of the data.

The leakage to state $|2\rangle$ is numerically evaluated in terms of the matrix elements of the full evolution operator as $\varepsilon_{\rm leak} = (|U_{02}|^2+ |U_{12}|^2)/2$. The detuning error can be estimated for the $X$ gate when $U_{\text{ideal}} =\sigma_x$.  In this case, the gate error without leakage can be estimated as $\varepsilon_{\rm det}=1-\mathcal{F} = 2(1-|U_{01}|^2)/3$. Both errors are estimated as a function of inverse gate time in Fig.~5b for the flux detuning of $\delta\varphi_{\rm{ext}}/2\pi = 3 \times 10^{-5}$. We find that the detuning error is negligible while the state $|2\rangle$ leakage is suppressed into the $10^{-4}$ range. However, at such an offset the coherence time already drops to about $40$~$\mu$s (see Fig.~4), which would bound the gate error at above $0.1\%$ for a gate time $t_{\pi} = 40$~ns. A smaller flux detuning would make both the leakage and the detuning errors negligible, while the higher detuning dynamics would be dominated by incoherent error.

We compare the measured gate fidelity with the numerical estimates in Fig.~5c using five different flux offset values and summing up the three possible errors. It appears that the data can be explained almost entirely by the incoherent error in the presence of a flux offset that varies on a long time scale in the range $\delta\varphi_{\rm{ext}}/2 \pi = (1-5) \times 10^{-5}$. We noticed the development of similar size offsets during long-time spectroscopy measurements. In general, such flux offsets are not uncommon in fluxoniums and they are usually mitigated by a more thorough magnetic shielding. Interestingly, for a flux offset below $\delta\varphi_{\rm{ext}}/2\pi< 10^{-5}$, the leakage to state $|2\rangle$ is negligible, in theory, even for relatively fast pulses ($t_{\pi} \approx 25$~ns). The ultimate flux-stability of integer fluxonium devices as well as the limitations on the lowest practical gate time will be investigated in future works.

\section{Discussion}\label{sec12}

Integer fluxonium qubit operates in zero magnetic field and in the same frequency range as transmons, it has a comparably high energy relaxation quality factor and gate fidelity, despite a much lower quality of the circuit dielectrics. Similarly to transmons, the degree of decoupling of integer fluxonium from the dielectric loss is controlled by the Cooper pair number matrix element $\langle 0|\hat{n}|1\rangle$. However, here the value of $\langle 0|\hat{n}|1\rangle$ can be controllably reduced by increasing the ratio $E_J/E_C$ almost independently of the qubit frequency. Also similarly to transmons, integer fluxonium remains unprotected against out-of-equilibrium decoherence mechanisms, such as quasiparticles in the superconductor or photons in the readout cavity.

The only apparent catch of the integer fluxonium design is the doublet nature of the qubit's excited state, which may result in a leakage of quantum information outside the computational subspace. Nevertheless, we experimentally demonstrated that even in a relatively extreme case of $\omega_{12}/2\pi = 11$ MHz (device D) the gate error remains under $0.1 \%$ with a gate time between 50-100 ns. This result was enabled by the parity selection rule, which inhibits the transition between states $|0\rangle$ and $|2\rangle$ exactly at $\varphi_{\textrm{ext}}=0$. Theoretical estimates of the leakage error are as low as $10^{-4}$ provided the flux bias is stable to about $\delta\varphi_{\rm{ext}}/2\pi\approx 10^{-5}$. The limits of flux stability will be investigated in a future work. \\

In this work, we left aside the effect of the finite device temperature. Indeed, even if the device temperature is as low as $10$ mK (the base temperature of dilution refrigerators), the state $|2\rangle$ in device D will have a nearly equal equilibrium occupation as state $|1\rangle$, since $k_B \times 10~{\rm{mK}}/h \approx 200~{\rm{MHz}} \gg \omega_{12}/2\pi = 11~{\rm{MHz}}$. A non-zero rate of thermal transitions between states $|1\rangle$ and $|2\rangle$ would thus introduce an additional leakage error. Reliably estimating this rate is difficult, as we know neither the effective temperature of the qubit environment nor the effective dielectric quality factor of this environment at such an unconventionally low frequency. Experimentally, the thermal leakage error is bound from above by the measured gate fidelity and hence it must be much lower than 0.1\%. In future experiments, the system can be studied at elevated temperatures and the population of state $|2\rangle$ can be reset by driving the transition $|2\rangle$- $|3\rangle$ accompanied by a fast relaxation from state $|3\rangle$ to state $|0\rangle$.

Two-qubit gates on a pair of capacitively coupled integer fluxonium devices can be implemented in the same way they were demonstrated on conventional fluxoniums~\cite{ficheux2021fast, xiong2022arbitrary}. In both cases, there is very little hybridization in the computational subspace due to the small values of the corresponding charge matrix elements. However, this is not the case for the non-computational subspace, and so connecting the two subspaces with a properly timed coherent pulse would produce a logical operation. In the case of conventional fluxonium, we demonstrated two-qubit operations around the frequency of the $|1\rangle$-$|2\rangle$ transition. Here, the plasmon corresponds to exciting the state $|3\rangle$ as can be seen in Fig.~1b. Therefore, a two-qubit gate can be achieved by driving the transition $|0\rangle$-$|3\rangle$  using the same optimization strategies explored in Refs.~\cite{ficheux2021fast, xiong2022arbitrary}. More detailed calculations of the two-qubit dynamics will be described elsewhere.

In closing, it is interesting to compare integer fluxonium to two seemingly very different superconducting devices. First is the charge qubit (Cooper pair box) in the limit $E_J \ll E_C$, operated at a zero charge offset. Integer fluxonium is an approximate dual to such an unbiased charge qubit, and it works much better because the $1/f$ flux noise amplitude is much smaller than the $1/f$ charge noise amplitude (when normalized to flux and charge quantum, respectively). Second is the ``soft $0-\pi$" qubit \cite{gyenis2021experimental}, with which integer fluxonium shares remarkably much in common. Both devices have a doublet qubit transition, leading to suppressed matrix elements of the flux and charge operators, combined with a first-order insensitivity to flux noise. It has been previously conjectured that achieving such a portfolio of properties requires a circuit with more than one degree of freedom~\cite{groszkowski2018coherence}. Indeed, the ``soft $0-\pi$" qubit is constructed by a relatively strong coupling of three modes: fluxonium-like, transmon-like, and harmonic oscillator. By contrast, integer fluxonium is described by a single degree of freedom. Understanding the profound difference between the two devices in terms of the origins of the noise decoupling could lead to new ideas for high-performance superconducting qubit designs.

\begin{acknowledgments}
We wish to acknowledge fruitful conversations with Chen Wang. This research was supported by the ARO HiPS (contract No. W911-NF18-1-0146) and GASP (contract No. W911-NF23-10093) programs.
\end{acknowledgments}

\appendix

\section{Fluxonium numerics}

The general fluxonium Hamiltonian is expressed as:
\begin{equation}
H = 4E_C \hat{n}^2 +\frac12 E_L \hat{\varphi}^2 - E_J \cos(\hat{\varphi} - \varphi_{\rm{ext}})
\label{daddy_app}
\end{equation}
The phase, $\hat\varphi$, and the Cooper pair number, $\hat n$, operators are expressed in the harmonic oscillator basis:
\begin{equation}
    \begin{aligned}
    \hat{\varphi} &=  
\frac{1}{\sqrt{2}} \bigg\{{\frac{8 E_C}{E_L}}\bigg\}^{1/4} ( {a}^{\dagger}+{a}) \ , \\
    \hat{n}  &= 
\frac{i}{\sqrt{2}} \bigg\{{\frac{E_L}{8 E_C}}\bigg\}^{1/4}  ( {a}^{\dagger}-{a}) 
    \end{aligned}
\end{equation}
 here $a,a^{\dagger}$ are the harmonic oscillator raising and lower operators each with a Hilbert space dimension exceeding fifty oscillator states. 

 After numerical diagonalization, the eigenvalues of the fluxonium Hamiltonian are returned; however, the eigenfunctions are the coefficients necessary to construct the fluxonium wavefunctions by the linear superposition of harmonic oscillator wavefunctions defined as:
\begin{equation}
\psi_k (\varphi) = \frac{1}{(2 \pi \varphi_{\rm zpf}^2)^{1/4}} \frac{1}{\sqrt{2^k k!}}H_k(\varphi/\varphi_{\rm zpf})e^{{- \varphi^2 /4\varphi_{\rm zpf}^2}} \ ,
\end{equation} 
where $k$ is the integer index of the harmonic oscillator state and $\varphi_{\rm zpf}= (8 E_C/E_L)^{1/4}$. Using these wavefunctions with the coefficients found by diagonalization we write $\Psi_j(\varphi) = \sum_{k=0}^{k=K} c_k \psi_k(\varphi)$, where $c_k$ is the $k$th component of the eigenvector for the $j$th eigenstate after diagonalizing and $\sum_k |c_k|^2 = 1$. 

\begin{figure}[t]
 \includegraphics[width=7.8cm]{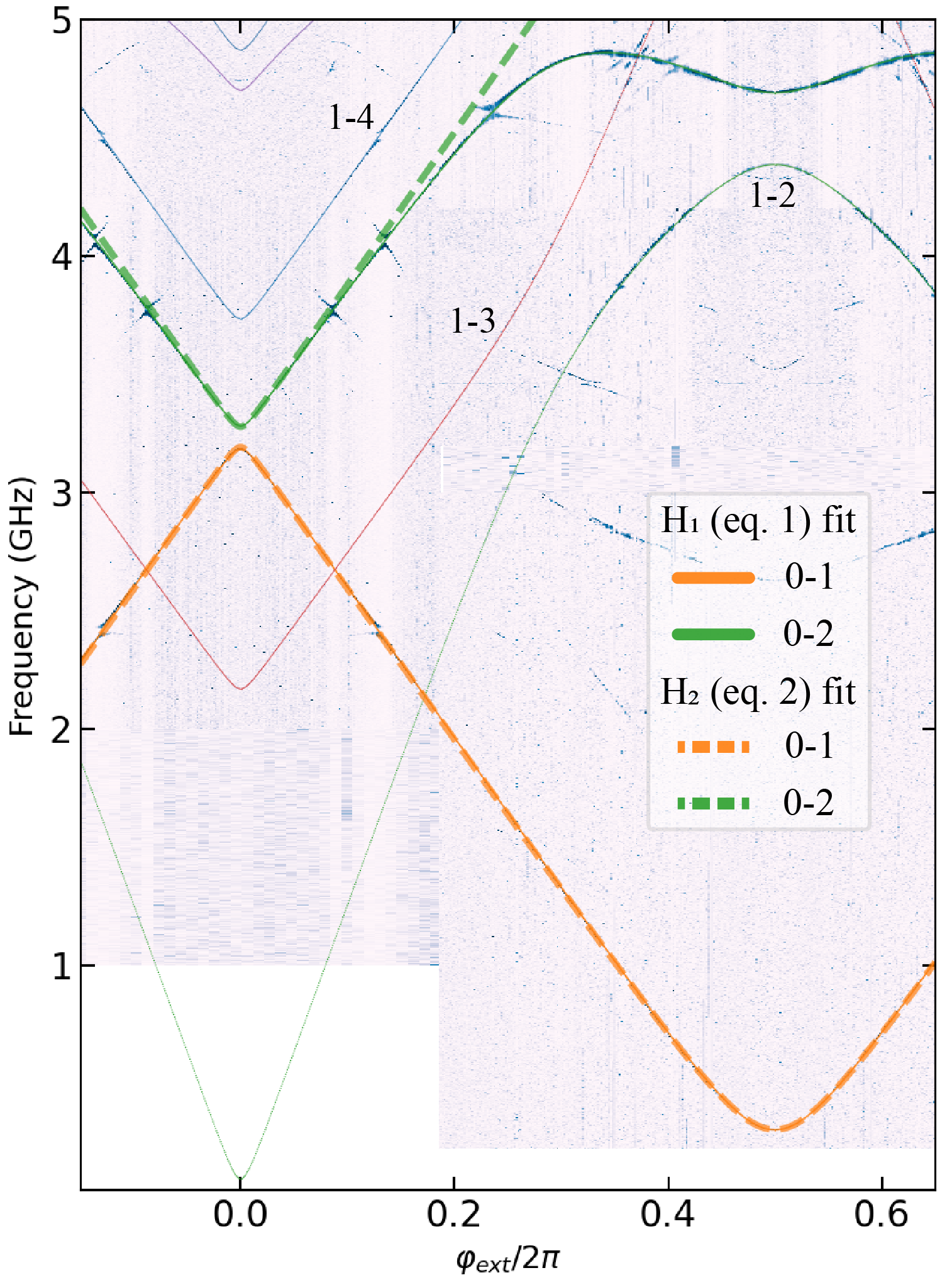}
 \caption{
 An example of two-tone spectroscopy of device A over the entire flux bias range superimposed with the fluxonium Hamiltonian model (Eq. 1, solid lines) and the effective Hamiltonian model (Eq. 2, dashed lines) fits. The thin lines correspond to transitions from the first excited state and are fit using Eq. 1.
 }
 \label{spectrumA}
 \end{figure}

\begin{figure}[t]
\includegraphics[width=8.5cm]{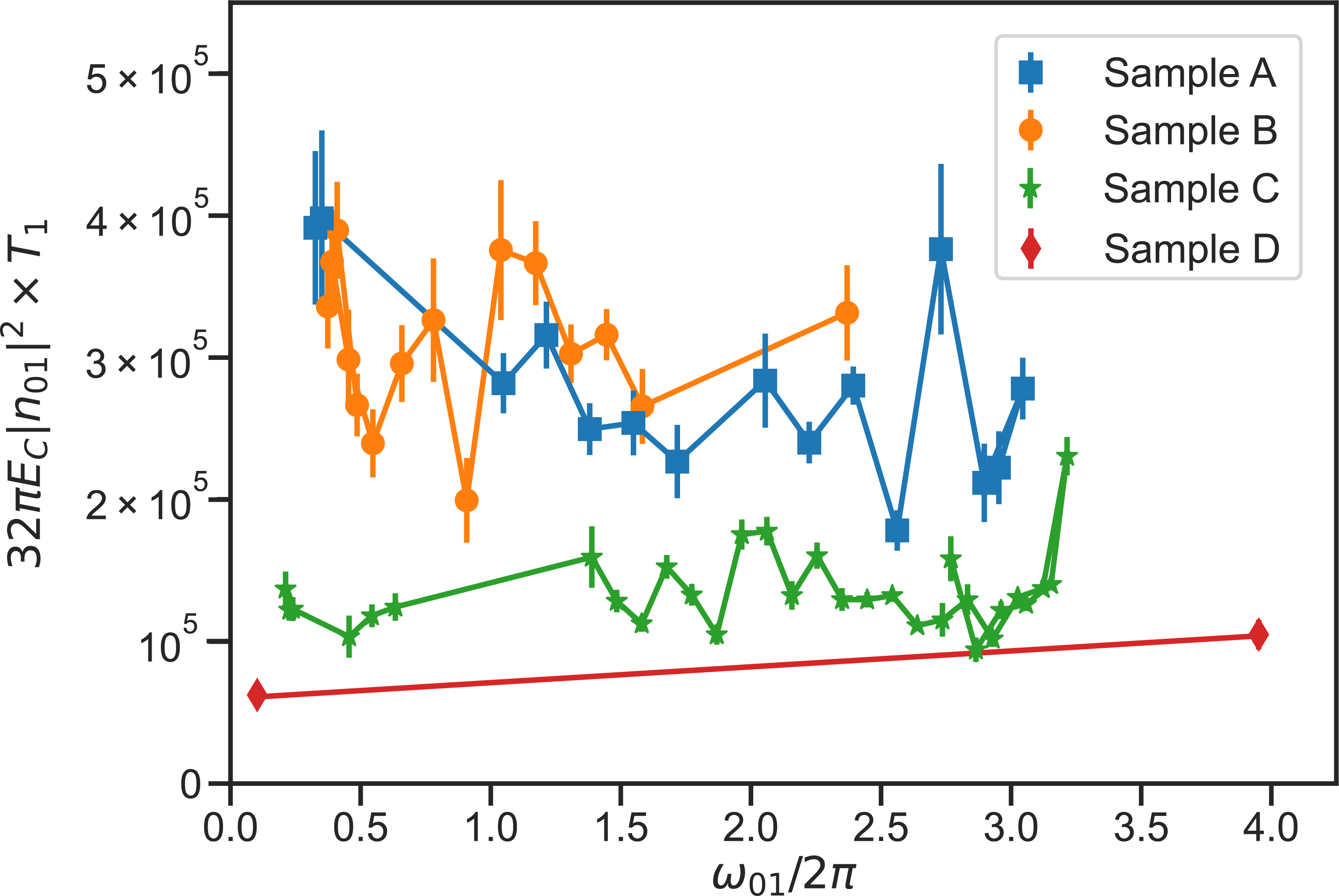}
\caption{The quantity $Q_{\textrm{diel}}= 32 \pi E_C |n_{01}|^2 \times T_1$ vs. the qubit frequency (see text arond Eq.~5) for four devices A,B,C, and D (see Table 1). The trend for frequency-independence over several octaves of frequency is consistent with the dielectric loss model. Data for devices A-C was compiled from Ref.~\cite{mencia2023ultra}. Device D happens to have the lowest value of $Q_{\rm{diel}}$, probably due to sub-optimal fabrication procedures.
}

\label{app:f2}
\end{figure}

 \section{Auxiliary data}

The experimental spectra for each device was fit using both the conventional fluxonium Hamiltonian Eq.~\eqref{daddy} and the discrete flux Hamiltonian Eq.~\eqref{baby}. Extracted parameters from both fit models for each device is summarized in Table 1. The spectrum of qubit device A is shown in Fig. 6. The fits to spectroscopy data show a near-perfect agreement of the fluxonium to the fluxon fit up to the vicinity of the plasmon (around 4 GHz in Fig. 6). The comparison of the fits verified that all devices $f_{01}^{\text{IFQ}}$ were indeed less than the plasmon mode frequency revealed by the first two transition's sharp external flux dependence in the vicinity of integer flux values. 

To further characterize qubits, $T_1$ was measured several times across half a flux quantum period for devices A-C; device D was only measured at both sweet spots. Per device, the multiple energy relaxation traces were first fit individually and were then binned together. The average decay constant per bin is the reported $T_1$ while the standard deviation per bin is the error. The binned statistics of energy relaxation times were then converted into an effective $Q_{\text{diel}}$ using Fermi's golden rule at zero temperature ($32 \pi E_C |n_{01}|^2 \times T_1$). The results are shown in Fig. 7 versus the qubit transition frequency and illustrate the effective $Q_{\textrm{diel}}$ does not depend explicitly on the qubit transition frequency. We note that incorporating a finite temperature to the lossy dielectric would only further increase the effective dielectric quality factor.

\newpage


%

\end{document}